\documentstyle[epsf]{mn}

\title[X-ray observations of B2 radio galaxies]{X-ray observations of low-power radio galaxies from the B2 catalogue}
\author[Canosa et al.]
       { C. M. Canosa$^{1}$, D. M. Worrall$^{1,2}$, M. J. Hardcastle$^{1}$ and M. Birkinshaw$^{1,2}$ \\
        $^{1}$Department of Physics, University of Bristol, Tyndall Avenue, Bristol, BS8 1TL\\
        $^{2}$Harvard-Smithsonian Center for Astrophysics, 60 Garden Street, Cambridge, MA 02138, U.S.A.}

\pubyear{1999}

\begin{document}

\maketitle

\label{firstpage}

\begin{abstract}

We present an analysis of X-ray data, taken with {\it ROSAT}, for a
well defined sample of low-power radio galaxies from the Bologna B2
catalogue. Where possible, the HRI has been used in order to take
advantage of the higher spatial resolution provided by this
instrument.  A variety of models are fitted to radial profiles in
order to separate the resolved and unresolved X-ray emission from the
galaxies. We demonstrate a strong, approximately linear, correlation
between the luminosities of the unresolved X-ray components and the
5-GHz luminosities of the radio cores in this sample. This suggests a
physical relationship between the soft X-ray emission of radio
galaxies and the jet-generated radio core emission. We infer a nuclear
jet-related origin for at least some of the X-ray emission.

\end{abstract}

\begin{keywords}
galaxies: active -- galaxies: clusters: general -- X-rays: galaxies.

\end{keywords}

\section{Introduction}

It is widely accepted that the radio emission in BL Lac objects is
dominated by synchrotron radiation from a relativistic jet pointed
towards the observer, thus explaining, among other things, the
superluminal velocities and rapid variability observed in several
objects. Such a favourable orientation to the observer should not be
common, implying the presence of a population of double radio sources
whose twin jets lie in the plane of the sky. Low-power radio galaxies
can be found which match BL Lac objects in extended radio power
(Wardle, Moore \& Angel 1984) and galaxy magnitude (Ulrich 1989).
There is evidence that the total soft X-ray luminosity in such galaxies is
correlated with the radio-core luminosity, implying a nuclear
jet-related origin for at least some of the X-ray emission (Fabbiano
et al. 1984). High spatial resolution X-ray measurements have further
strengthened this argument by separating point-like emission from hot
X-ray emitting atmospheres (Worrall \& Birkinshaw 1994; Edge \&
R\"ottgering 1995; Feretti et al. 1995; Worrall 1997; Hardcastle \&
Worrall 1999). What has been lacking is high-resolution X-ray
observations of a large unbiased sample of low-power radio galaxies
with which to investigate the association of unresolved X-ray emission with the
nuclear radio jet. 

The B2 bright sample of radio galaxies are Bologna Catalogue 408-MHz
radio sources identified with elliptical galaxies brighter than
m$\rm_{Zwicky}$=15.4 (Colla et al. 1975, Ulrich 1989). The radio
survey occupies 0.84 steradian, and is complete at 408 MHz down to 0.2
Jy for (B1950) declinations between 29 and 34 degrees, 0.25 Jy for
declinations between 24 and 29.5 degrees and between 34 and 40
degrees, and 0.5 Jy between 21.4 and 24 degrees (Colla et al. 1970,
1975; Ulrich 1989), although none of the bright sample radio galaxies
lies in the last declination range. The sample comprises 50 galaxies,
of which 47 are at $z \leq 0.065$ (see Table
\ref{sample}). The high Galactic latitudes imply a relatively small
Galactic neutral hydrogen column density and small resultant X-ray
absorption. The B2 sample has been shown to be well matched with
radio-selected BL Lac objects in their extended radio properties and
galaxy magnitudes (Ulrich 1989). In this paper we present {\it ROSAT}
X-ray measurements of 40 galaxies which constitute an unbiased
subsample of the 47 galaxies at $z \leq 0.065$. The data were taken
from our pointed observations or from data in the {\it ROSAT} public
archives.

Section 2 discusses the sample of galaxies observed with $\it ROSAT$ and the general properties of the X-ray data. Details
of our analysis and notes on some of the sources are in section
3. Section 4 describes X-ray -- radio comparisons. Section 5 contains
our conclusions.

A Friedmann cosmological model with $H_{0} =50$ km s$^{-1}$ Mpc$^{-1}$,
$q_{0} =0$ is used throughout this paper.

\section{X-ray data}

Table \ref{sample} lists the sample of 50 B2 radio sources associated
with elliptical galaxies; we have not included B2 1101+38 and B2 1652+39 as these are the well known BL Lac objects Mkn 421 and Mkn 501. We tabulate the pointed observations taken with {\it
ROSAT} using the instrument with the highest spatial resolution, the
High Resolution Imager (HRI; David et al. 1997), or, if no HRI
observations were made, the Position Sensitive Proportional Counter
(PSPC; Tr\"umper 1983; Pfeffermann et al. 1987).  Where multiple
observations exist, we present results for the longest on-axis
viewings, concentrating where possible on HRI data. Most observations
result in a detection. In some cases where good data exist from the
HRI and PSPC, both were analysed and compared (this was the case for
B2 0055+30, 0120+33, 0149+35, 1122+39; 1217+29 and 2335+26). Within
the same ROR, some HRI observations were split into observing periods
about 6 months or a year apart. Where this was the case, the data have
been merged after the individual `observations' were checked for
anomalies. In many cases resolved X-ray emission (measured better with
the PSPC) is seen in addition to point-like emission. Complementary
work discussing the extended X-ray emission and X-ray spectra of B2
radio galaxies based on PSPC observations can be found in Worrall $\&$ Birkinshaw (in preparation). Our treatment of the resolved emission in this
paper is restricted to modelling it sufficiently well to determine a
best estimate for the contribution from central unresolved emission.

\begin{figure*}
\label{blocks}
\begin{center}
\leavevmode
\epsfxsize 15cm
\epsfbox{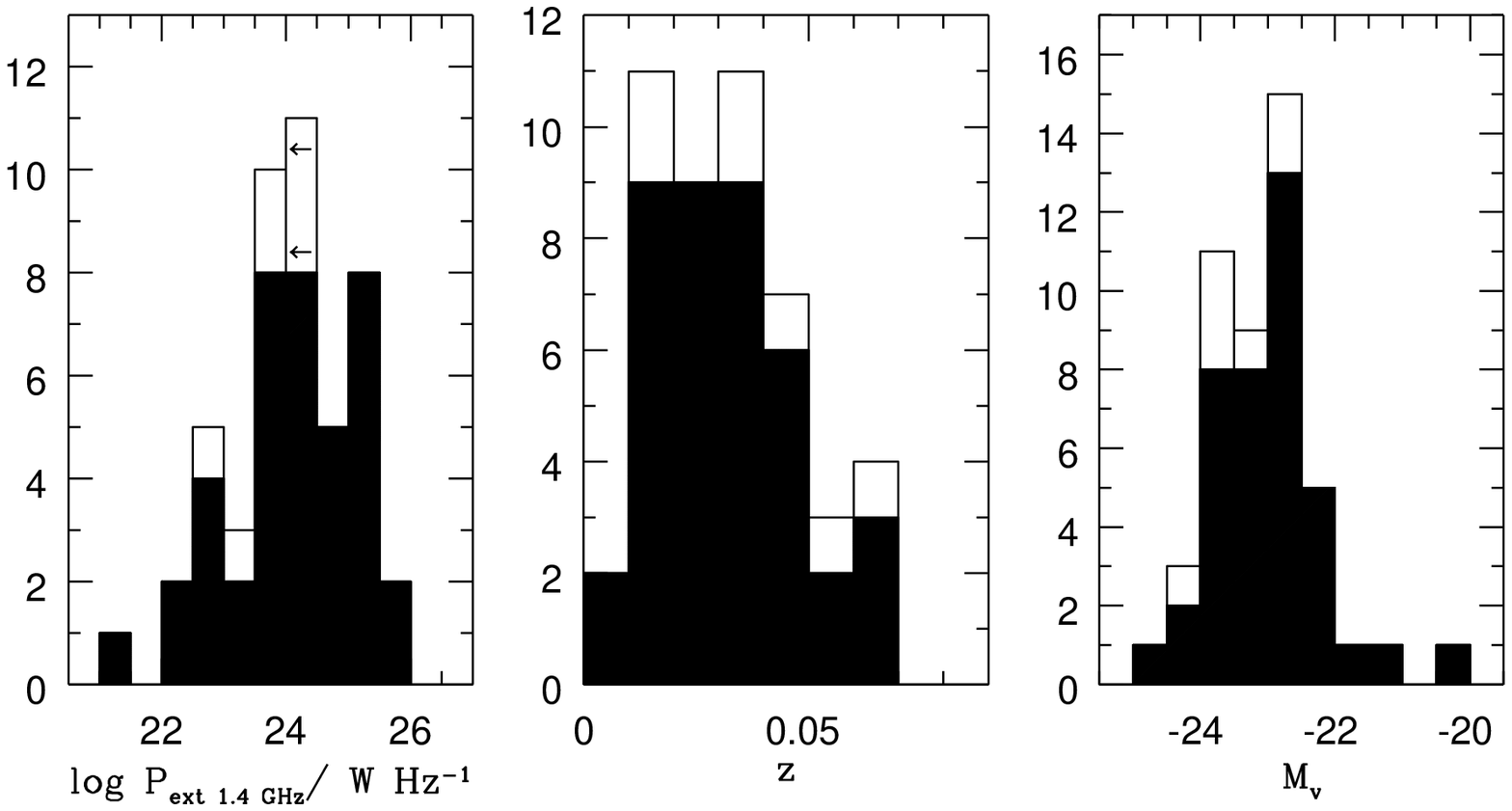}
\end{center}
\caption{Histograms of extended radio power at 1.4 GHz, redshift and
galaxy V-band magnitude for the 47 B2 $z\leq 0.065$ sources (Ulrich
1989). Solid shading shows the 40 sources for which we have {\it
ROSAT} X-ray data.}
\end{figure*}

40 of the full sample of 50 B2 radio galaxies (and the 47 at $\it z <$0.065) listed in Table \ref{sample} were observed with $\it ROSAT$: the demise of the satellite in early 1999 prevented observation of the remaining 7 objects with $\it z <$0.065. No known bias was introduced into the set of 40 objects for which we have data by the process of prioritizing sources for $\it ROSAT$ observation. This is illustrated in Fig. \ref{blocks}, which shows histograms of 1.4-GHz extended radio power ($P
\rm_{ext}$), redshift, and absolute visual magnitude ($M\rm_{v}$) for
the 47 galaxies at $z\leq 0.065$. $P\rm_{ext}$ and $M\rm_{v}$ are
indicators of isotropic unbeamed emission used to support the
association of these galaxies with the hosts of BL Lac objects. A Kolmogorov-Smirnov test finds no significant ($>$90 per cent confidence) difference between the distributions of sources for which we have $\it ROSAT$ data, and those for which we do not, in any of the quantities  $P \rm_{ext}$, $\it z$, or $M\rm_{v}$.

The {\it ROSAT} HRI has a roughly square field of view with 38 arcmin
on a side.  A functional form for the azimuthally averaged point
spread function (PSF) can be found in David et al. (1997). The core
of the HRI radial profile of a point-like source may be wider than the
nominal PSF, and ellipsoidal images are sometimes seen, due to a
blurring attributable to residual errors in the aspect correction. The
major axes of such ellipsoidal images are not aligned with the
satellite wobble direction (the wobble is employed to ensure that no
sources are imaged only in hot pixels in the HRI or hidden behind the
PSPC window-support structure) and depend unpredictably on the day of
observation and therefore the satellite roll angle. The asymmetry
is strongest between 5 arcsec and 10 arcsec from the centroid of the
image. The PSF for the HRI begins to be noticeably influenced by the
off-axis blur of the X-ray telescope at $\ga$7 arcmin off-axis. All
HRI observations discussed here are essentially on axis.

The {\it ROSAT} PSPC has a circular field of view, of diameter 2
degrees, and the PSF has a FWHM of about 30 arcsec at the center of
the field, degrading only marginally out to off-axis angles of about
20 arcmin. At larger radii the mirror blur dominates the spatial
resolution, and at 40 arcmin off axis the FWHM of the PSF is about 100
arcsec. Only one of the 40 sources discussed here (B2 0722+30) is
significantly affected by mirror blur, as all others were observed in
the central part of the field of view.

We used the Post Reduction Off-Line Software (PROS; Worrall et
al. 1992) to generate radial profiles of the X-ray data. Background
was taken from a source-centered annulus of radii given in Table
\ref{counts}. Where the radial profile has been used to probe the
extended emission, the contribution from extended-emission models to
the background region is taken into account (Worrall \& Birkinshaw
1994).

We excluded confusing sources, defined as those separated from the
target by $\ga 15$ arcsec (HRI) and $\ga 30$ arcsec (PSPC), showing up
at $\sim 3\sigma$ above the background level and overlapping the
on-source or background regions (or being slightly beyond, but still
affecting the background due to the broad wings of the PSF). Optical
images (e.g the Palomar Sky Survey plates, digitized by the Space
Telescope Science Institute) and radio maps were overlaid on the X-ray
images, to check the identity of each target source and any
neighbouring X-ray sources, and to help to classify and
limit further the effects of confusing sources.

Our luminosity determination for unresolved emission assumes a
power-law spectrum with an energy index $\alpha$ of 0.8 ($f _{\nu} \propto
\nu^{- \alpha}$) modified by Galactic absorption. Variations in
spectral form affect the luminosity, but this is normally a small
error compared with statistical uncertainties.

\section{Radial profiles and nuclear X-ray emission}

We analyse the structure of an X-ray source by extracting a radial
profile and fitting various models convolved with the energy-weighted
PSF of the detector in question [the nominal PSF for the HRI (David et
al. 1997) and the PSF for the PSPC (Belloni, Hasinger
\& Izzo 1994)].

As well as point-source models, we have fitted our radial profiles
with $\beta$ models (Sarazin 1986) which describe gas in hydrostatic
equilibrium. The values used for $\beta$ were 0.3, 0.35, 0.4, 0.5, 0.67, 0.75
and 0.9. For each source the models used are (a) point source, (b)
$\beta$ model and (c) point source + $\beta$ model. Model parameters
for sources where an extended component is detected are given in Table
\ref{models}. In most cases we find that for observations with enough
counts a significantly better fit to the data is achieved by fitting
the composite model (c) rather than either (a) or (b)
individually (see Table \ref{models}).

Residual errors in aspect correction affect the HRI PSF and must be
taken into account when evaluating the contribution from a point
source [see Worrall et al. (1999), where it was found that a point
source can appear more like a $\beta$ model with a core radius of up
to about 5 arcsec (dependent on $\beta$ assumed) because of the aspect
smearing]. Of the sources in this paper only 1833+32 has high enough
count rate to perform the dewobbling procedure described by Harris et
al. (1998); we find no substantial difference in the best-fit
parameters after dewobbling. For observations where the data show
extension on small scales, it is difficult to decide whether the
source is really point like or indeed slightly extended. In some
cases, the fit to the point+$\beta$ model has a substantially lower
$\chi^2$ than the beta-model fit alone, and here we regard the
point-like component as well measured. Where $\chi^{2}$ remains
unchanged for these two models, the total counts have been taken as an
upper limit on the point-source counts. A literature search into the
environments of the sources and a cross-correlation of the HRI and
PSPC results, where possible, helps us further validate our HRI
findings and place limits on the likelihood of the source being
primarily point-like.

In the 7 cases where there are not enough counts to perform adequate
radial-profile fitting , the total counts are taken as an upper limit
to the contribution of point-like emission. For non-detections a
3$\sigma$ upper limit, derived by applying Poisson statistics to a 5
by 5 arcsec detection cell for the HRI on-axis observations (3 cases),
and a 120 by 120 arcsec detection cell for the PSPC observation of the
off-axis source B2 0722+30, centred on the position of the radio
and optical core, is taken as an upper limit on both the total and the
unresolved X-ray emission.

Table \ref{models} gives results for the 16 sources with enough counts
to allow radial-profile model fitting. Table \ref{counts} presents the
net counts within a circle of specified radius and, for the sources
with enough counts to allow radial profiling (consisting of a minimum of $\sim$70 counts over 3 data bins), the point-source
contribution. There is a wide range in the ratio of unresolved to
resolved counts. X-ray core flux and luminosity densities calculated
from the unresolved count rates, and radio core flux and luminosity
densities taken from the literature are given in Table \ref{lums}.

\subsection{Notes on individual sources}

Where analysis of sample sources is present in the literature,
comparisons have been made with the results presented here. Results
for B2 0055+26 are taken from Worrall, Birkinshaw $\&$ Cameron
(1995). B2 0326+39, 1040+31 and 1855+37 are discussed in detail in
Worrall $\&$ Birkinshaw (in preparation). The HRI results for B2 2229+39 are
consistent with the findings of Hardcastle, Worrall $\&$ Birkinshaw
(1998), for a PSPC observation of the source.

\noindent {\it B2 0120+33} \\ Identified with the galaxy NGC 507, the source
lies in Zwicky cluster 0107+3212 and is one of the brightest
galaxies in a very dense region. Extended X-ray emission is seen out
to a radius of at least 16 arcmin, and there is evidence for the
presence of a cooling flow and possible undetected cooling clumps
distributed at large radii (Kim \& Fabbiano 1995). B2 0120+33 has a
steep radio spectrum and weak core. It may be a source with
particularly weak jets, or possibly a remnant of a radio galaxy whose
nuclear engine is almost inactive and whose luminosity has decreased
due to synchrotron or adiabatic losses (Fanti et al. 1987).

The HRI map shows the central region of this source to be
asymmetrical, with a large extended emission region to the
SW. Radial-profile fitting of the innermost parts of this galaxy with
point, $\beta$, and point+$\beta$ models, show that a good fit to
the data is achieved by using a single $\beta$ model with
$\beta$=0.67, which gives a core radius of 4$\arcsec$. The
$\beta$+point model gives a marginally better fit, but the additional
component is not significant on an F-test at the 90$ \%$ confidence
level. Therefore, we have taken the total counts from the inner
regions of this source as an upper limit on any unresolved emission
present, keeping in mind that this may also include a cooling-flow
contribution (this also holds for other sources with possible
unresolved cooling flows such as B2 0149+35, 1346+26 and 1626+39). 

\noindent {\it B2 0149+35} \\ B2 0149+35 is identified with NGC 708 and is
associated with the brightest galaxy in the cluster Abell 262. Braine
\& Dupraz (1994) suggest that it contains a cooling flow which may
contribute excess central X-rays, and this may explain why B2 0149+35
has a higher point-like X-ray luminosity and flux than expected based
on other sample members. It is not possible to separate spatially a
cooling-flow contribution from unresolved X-ray emission using the
PSPC observation, and the asymmetry of the source makes the extraction
of a radial profile difficult.  \\ 
The HRI observation is split up into 2 OBIs (Observation Intervals),
one of which shows a barred N-S structure. Each OBI was individually
analysed by taking close-in source regions, and this gives 
results which are consistent with the PSPC data from the inner region of B2 0149+35. In the
longer, and more reliable, of the two OBIs (12.7ks), 207 counts were
detected. The point-like contribution to the net emission from this
source however, is not significant at the 95 per cent level when an F-test is
performed. The detected counts are therefore taken as an upper
limit on the point model emission. The shorter OBI was not used.

\noindent{\it B2 0207+38} \\ This source is described as being more similar to
an S0 or to a spiral galaxy than an elliptical (Parma et al. 1986). It
has also been called a post-eruptive Sa (Zwicky et al., 1968) or a
distorted Sa (de Vaucouleurs, de Vaucouleurs \& Corwin 1975). The
radio structure is disc-like and there is no sign of either a radio
core, or of jets or radio lobes (Parma et al. 1986). It is probably a
starburst, like B2 1318+34. There are no {\it ROSAT} X-ray data
for this source.

\noindent{\it B2 0836+29A} \\ This object (4C 29.30), this object has been often confused in the literature with the cD galaxy B2 0836+29 at $\it z$=0.079, which is the brightest galaxy in Abell 690.

\noindent{\it B2 0924+30} \\ B2 0924+30 appears to be a remnant radio
galaxy whose nuclear engine is inactive (Fanti et al., 1987; Cordey
1987; Giovannini et al. 1988). It is the brightest member in a Zwicky
cluster (Ekers et al. 1981), and the X-ray data suggest extended X-ray
emission, although the detection is of marginal significance. The
relatively high X-ray emission for a source with no detectable core
radio emission may therefore be due to the extended gas in the
cluster. We have taken the detected emission as an upper limit on
possible point source emission.

\noindent{\it B2 1122+39} \\ Analysis of this source, both in the PSPC and in the HRI, shows that $\sim$3 per cent of the total emission is contributed by an unresolved source. This is consistent with the findings of Massaglia et al. (1996) who find a contribution from a point source of $<$6 per cent.

\noindent{\it B2 1217+29} \\ PSPC and HRI analysis of this source are consistent. A $\beta$+point model fits the data better than a $\beta$ model alone (see Table \ref{models}).

\noindent{\it B2 1254+27} \\ There is a large discrepancy between the positions
given for this object in the NASA Extragalactic Database (NED) and the
SIMBAD Astronomical Database. This is because the radio source has, in
some cases, been incorrectly associated with the galaxy NGC 4819
rather than the true host galaxy for the radio emission which is NGC
4839.

NGC 4839 is classified confusingly as morphological type S0 (Eskridge
\& Pooge 1991), E/S0 (Jorgensen, Franx \& Kjaegaard 1992) or as a cD
(Gonzalez-Serrano, Carballo \& Perez-Fournon 1993; Fisher, Illingworth
\& Franx 1995). Andreon et al. (1996) also mention that the low
average surface brightness suggests that this galaxy is dominated by
an extended disc. The X-ray map of the source shows extended large scale emission to the SW [described by Dow \& White (1995), as being in the
process of interacting with the intracluster medium of the main (Coma)
cluster]. This goes beyond the size of the optical galaxy and has been excluded here so as not to affect the background emission. About 88 per cent of the net counts arise from a point-like emission component.

\noindent{\it B2 1257+28} \\ The region of enhanced X-ray emission in B2 1257+28 in the Coma cluster is substantially smaller than the size of the optical
galaxy. 
Small on-source (12 arcsec radius) and background (15-22.5
arcsec) source-centered circles were used in order to verify the
contribution from unresolved X-ray emission given by our best-fit model.

\noindent{\it B2 1317+33} \\ B2 1317+33 (NGC 5098A) has a companion galaxy (NGC
5098B) at a distance of $\sim 40 $ arcsec. We have checked that the
X-ray and radio source come from NGC 5098A by overlaying the radio,
optical and X-ray maps.

\noindent{\it B2 1318+34} \\ B2 1318+34 is a classic merger-induced starburst,
whose total radio flux can be attributed to starburst activity rather
than an active nucleus (Condon, Huang \& Yin 1991).

\noindent{\it B2 1346+26} \\ The source is a cD galaxy in Abell 1795,
identified with 4C 26.42. It contains a central cooling-flow
component, as discussed in Fabian et al. (1994). $\it HST$ WFPC2 images
of the core of this cooling flow are presented in Pinkney et
al. (1996).

Our analysis of the HRI data detects a central point source which is
significant on an F-test at the 95 per cent confidence level (see
Table \ref{models}). About 1 per cent of the total counts lie in this
point-like component (see Table \ref{counts}).

\noindent{\it B2 1422+26} \\ B2 1422+26 is not radially symmetric in the
X-ray. An off-axis X-ray source in the same field of view gives a good
fit to the nominal PSF, and so we can rule out the possibility that
the X-ray extension seen in B2 1422+26 is due to the {\it ROSAT}
aspect correction problem.  The possible detection of a point-like
component is not significant on an F-test at the 95 per cent level
(although it passes at the 90 per cent level). We have nevertheless
taken the point counts calculated from these model fits as our best
estimate of the central emission, though the errors are large.

\noindent{\it B2 1615+35} \\ HRI analysis is consistent with Feretti et al. (1995). The X-ray emission is largely point like (see Table \ref{counts}), with $\sim$60 per cent of the net counts coming from the point source.

\noindent{\it B2 1621+38} \\ B2 1621+38 was analysed by Feretti et al. (1995), who found a point-source contribution of $\leq$50 per cent of the total X-ray flux. Our analysis is consistent with this result; we find the point-source contribution to be
18$\pm$4 per cent.

\noindent{\it B2 1626+39} \\ B2 1626+39 lies in a cluster (A2199), with a prototypical cooling flow. Owen \& Eilek (1998) conclude that the radio source is
relatively young and has been disrupted by the surrounding gas. The
{\it ROSAT} HRI data set for this source consists of 2 OBIs roughly 7
months apart. Only in the second observation does the source appear
extended, with two adjacent peaks. We have taken this to be due to
errors in the aspect correction or processing effects and therefore
have used only the first OBI in our analysis.

\noindent{\it B2 1833+32} \\ B2 1833+32 is an FRII radio galaxy 
(Laing, Riley \& Longair 1983; Black et al. 1992) with broad emission
lines (Osterbrock, Koski \& Phillips 1975; Tadhunter, Perez \& Fosbury
1986; Kaastra, Kunieda \& Awaki 1991). Its higher than expected X-ray
flux, as compared with the core radio strength, may arise from
emission in the central accretion disc around the active nucleus, seen
due to an advantageous viewing angle as indicated by the broad
emission lines.

\noindent{\it B2 2236+35} \\ This source has a double symmetric radio jet embedded in a low surface-brightness region. The two extended lobes are similar in strength and size (Morganti et al. 1987). The X-ray emission at radii greater than about 20 arcsec seems aligned with the radio jets in this source. Model fitting shows $\sim$25 per cent of the total counts to be in the point source.

\section{X-ray -- radio comparisons}

In Table \ref{counts} we list the net counts detected for each source
within a specified radius and our best estimate of the contribution
from unresolved emission derived as described above.  1-keV luminosity
densities and broad-band soft X-ray luminosities calculated from the
values for unresolved emission are given in Table \ref{lums}, along
with the radio core flux density and luminosity density. In Fig. 2 we
show a logarithmic plot of radio core luminosity density against the unresolved
core X-ray emission; the corresponding flux density-flux density plot
appears in Fig. 3. 

Both the logarithmic X-ray and radio flux densities and the corresponding
luminosities are correlated at the $>99.99$ per cent significance
level on a modified Kendall's $\tau$-test which takes upper limits
into account, as implemented in ASURV (Lavalley, Isobe \& Feigelson
1992). The flux-flux correlation gives us confidence that the
luminosity-luminosity relationship is not an artificially-introduced
redshift effect.

To determine the slope of the core flux-flux and luminosity-luminosity
plots, a generalised version of the Theil-Sen estimator was used as
presented in Akritas, Murphy $\&$ LaValley (1995). This takes into
account the nature of the upper limits by assuming that the individual
points are all part of the same parent population. For more details of
this analysis and its advantages over the more commonly used
survival-analysis method, see Hardcastle \& Worrall (1999), where it
is explained how using the bisector of two regression lines provides a
more robust estimate of the slope. This method however does not give a value for the intercept of the best-fit line. To determine the best-fit line
plotted on Fig. 2, a regression based on the bisector of slopes
determined by the Schmitt algorithm as implemented in ASURV was used,
because it does allow us to determine an intercept.

For the whole sample, the Theil-Sen luminosity-luminosity slope is
1.05 with 90 per cent confidence limits 0.86-1.28.
There are however, a few galaxies that should be removed. These are
the starburst object B2 1318+34 (which is not member of a strict AGN
sample), and the FRII B2 1833+32 (which is a broad-line galaxy). For
comparison purposes we have retained these objects on Figs. 2 \&
3. With the omission of these two sources, we find a Theil-Sen slope of 0.96 for
the luminosity-luminosity relation, with a 90 per cent confidence
range (derived from simulation) of 0.78 -- 1.21. The median
logarithmic dispersion about the regression line is $\sim 0.2$.
Schmitt regression analysis of the slope gives consistent results, with a slope of 1.15 and a 90 per cent confidence limit of 0.92 -- 1.39. Normalisation of the Schmitt slope at a radio luminosity of 10$^{22}$W Hz$^{-1}$sr$^{-1}$, gives the best-fit normalisation value of  15.55 (i.e. the predicted X-ray flux at that radio luminosity is 10$^{15.55}$W Hz$^{-1}$sr$^{-1}$. The 90 per cent confidence range on this is 15.2 to 15.9.

From combining statistical methods, the best overall estimate and 90 per cent confidence uncertainty for the core X-ray--radio luminosity relation for low-power radio galaxies is given by:

	log(l$\rm_{x}$) = (0.96$^{+0.25}_{-0.18}$) log (l$\rm_{r}$/10$^{22}$) + (15.55 $\pm$ 0.35) \\

A couple of sources lie significantly away from the regression line
(B2 1254+27 and B2 1257+28). Both lie in the Coma cluster and may be
contaminated by cluster emission.

B2 0120+33 has been classified as a possible remnant of a radio galaxy
whose nuclear engine is inactive (Fanti et al. 1987), and is shown as
an upper limit above the expected value based on the X-ray correlation
for other sample members. Its high total X-ray flux may be due to the
cooling flow seen by Kim \& Fabbiano (1995) (see Section 3 of this
paper) and makes the extraction of the core flux difficult. B2 0924+30
is also a relic source. B2 0149+35 contains a cooling flow which may
contribute excess central X-rays (Braine \& Dupraz 1994).

The correlation shown in Figs. 2 \& 3 suggests a physical relationship
between the soft X-ray emission of radio galaxies and the
jet-generated radio core emission. Correlations between the total
X-ray emission and the radio core emission have been seen in the $\it
 Einstein$  $\it Observatory$ data (e.g Fabbiano et al. 1984), but did
not have the spatial resolution necessary to separate point and
extended components.  Our $\it ROSAT$ analysis and the decomposition
of the X-ray emission into resolved and unresolved components, now
shows that the nuclear X-ray emission is strongly correlated with the
nuclear radio-core emission. This favours models which imply a nuclear
jet-related origin for at least some of the X-ray emission.

\section{Conclusions}

Radial profiling and model fitting of {\it ROSAT} data, primarily from
the HRI, have allowed us to separate point-like contributions from the
overall X-ray emission in low power radio galaxies from a well
defined, nearby sample. We find flux-flux and
luminosity-luminosity core X-ray/radio correlations for such sources,
with slopes that are consistent with unity. This suggests a physical
relationship between the soft X-ray emission of radio galaxies and the
jet-generated radio core emission, with the clear implication that at
least some of the X-ray emission is related to the nuclear radio
jet. In future work we will estimate X-ray beaming parameters under the
assumption that radio galaxies are the parent population of BL Lac objects.

\section*{Acknowledgments}

This research has made use of the NASA/IPAC Extragalactic Database
(NED) which is operated by the Jet Propulsion Laboratory, California
Institute of Technology, under contract with the National Aeronautics
and Space Administration. Support from NASA grant NAG 5-1882 is
gratefully acknowledged. We thank the referee for useful comments.

\clearpage

\clearpage

\begin{table*}
\caption{Data on the 50 B2 sample galaxies. RA and DEC are given for
the radio core where possible, and the optical identification
otherwise (optical positions are labelled with a *). References are as
follows: 1, Fanti et al. (1987), 2, Worrall et al. (1995); 3, Ekers
(1978); 4, Fanti et al. (1982); 5, Bridle et al. (1991); 6, Measured
from radio map in Laing (1996); 7, Johnston et al. (1995), 8, Measured
from radio maps supplied by L. Rudnick; 9, Venturi et al. (1995); 10,
Stocke \& Burns (1987); 11, Fanti et al. (1973); 12, Schillizzi et
al. (1983), 13, Kim (1994); 14, Bridle et al. (1981); 15, Giovannini et
al. (1988); 16, Estimated from map of Venturi et al. (1993). Galactic
absorbing column density $N_{\rm H}$ is taken from Stark et
al. (1992). Redshifts are from Colla et al. (1975). $M_{\rm v}$ is the
absolute visual magnitude of the host galaxy from Ulrich (1989). The
ROR is the {\it ROSAT} observational request number, with `rh'
referring to HRI observations and `rp' to PSPC observations. Offset
refers to the off-axis angle, and is shown only for those sources with
offsets $\geq$6 arcmin. A dash in the last three columns indicates no
observation.}
\label{sample}
\begin{tabular}{@{}lllllllllll@{}}
   B2     & Other&RA  & DEC  &  Ref. &$N_{\rm H}$ & $z$ &  $M_{\rm v}$ & ROR & Livetime  & Offset \\
name    &name & (J2000) & (J2000) & & ($10^{20}$ & &  & & (s) & (arcmin)\\
& & & & &  cm$^{-2}$)& &  & &\\
\hline
0034+25 &&00 37 5.50 &+25 41 56.4 &1 &3.84 &0.0321 &$-$22.95 &--&--&--\\
0055+30 &NGC315&00 57 48 &+30 21 08 &16 &5.77 &0.0167 &$-$23.68 &rh701308n00 & 14821&--\\
0055+26 &NGC326&00 58 22.63 &+26 51 58.8 &2 &5.47 &0.0472 &$-$23.78 &rp700884n00 & 20356&--\\
0104+32 &3C31&01 07 24.9 &+32 24 45.00 &6 &5.53 &0.0169 &$-$22.60 &rh600496n00 & 24815&--\\
0116+31 &4C31.04&01 19 35.0 &+32 10 50 &7 &5.82 &0.0592 &$-$23.50 &--&--&--\\
0120+33 &NGC507&01 23 39.99 &+33 15 21 &1 &5.41 &0.0164 &$-$23.14 &rh600680n00 & 27992&--\\
0149+35 &NGC708&01 52 46.46 &+36 09 06.6 &1 &5.30 &0.0160 &$-$21.70 &rh800870n00 & 12715&--\\
0206+35 &4C35.03&02 09 38.55 &+35 47 51.1 &1 &5.90 &0.0375 &$-$23.34 &rh701564n00 & 21408&--\\
0207+38 &NGC828&02 10 09.7 &+39 11 30 $^*$&1 &5.36 &0.0181 &$-$23.58 &--&--&--\\
0222+36 &&02 25 27.34 &+37 10 27.8 &1 &5.19 &0.0327 &$-$22.80 &rh704037n00 & 28701&--\\
0258+35 &NGC1167&03 01 42.4 &+35 12 20 &1 &9.26 &0.016 &$-$22.30 &rh704039n00 & 18992&--\\
0326+39 &&03 29 23.9 &+39 47 32 &5 &14.21 &0.0243 &$-$22.50 &rp701442n00 & 18931&--\\
0331+39 &4C39.12&03 34 18.42 &+39 21 24.3 &1 &14.60 &0.0202 &$-$22.88 &rh701832n00 & 25993&--\\
0648+27 &&06 52 02.51 &+27 27 39 &1 &12.23 &0.0409 &$-$24.03 &--&--&--\\
0722+30 &&07 25 37.26 &+29 57 14.8 &1 &6.48 &0.0191 &$-$21.28 &rp200450n00 & 11362&28.0\\
0755+37 &NGC2484&07 58 28.18 &+37 47 12.5 &1 &5.02 &0.0413 &$-$23.50 &rh701309n00 & 19509&--\\
0800+24 &&08 03 16.6 &+24 40 34 &10 &4.30 &0.0433 &$-$22.98 &rh702077n00 & 29936&--\\
0836+29A &4C29.30&08 40 02.35 &+29 49 02.2 &1 &4.11 &0.0650 &$-$23.88 &rh800159 & 19036&--\\
0844+31 &4C31.32&08 47 59.02 &+31 47 09 &1 &3.45 &0.0675 &$-$24.15 &--&--&--\\
0915+32 &&09 18 59.4 &+31 51 40.4 &1 &1.86 &0.0620 &$-$23.98 &--&--&--\\
0924+30 &&09 28 52.8 &+29 59 07 $^*$&11 &1.88 &0.0266 &$-$22.88 &rh701831 & 28722&--\\
1040+31 &4C29.41&10 43 19 &+31 31 01 &1 &1.82 &0.036 &$-$22.58 &rp700883n00 & 21248&--\\
1102+30 &&11 05 22.81 &+30 09 41.5 &1 &1.98 &0.0720 &$-$24.08 &--&--&--\\
1108+27 &NGC3563&11 11 25.20 &+26 57 48.8 &1 &1.66 &0.0331 &$-$23.28 &--&--&--\\
1113+29 &A1213&11 16 34.5 &+29 15 16 &1 &1.57 &0.0489 &$-$23.70 &rp800167 &  6547& 6.5\\
1122+39 &NGC3665&11 24 43.5 &+38 45 47 &1 &2.08 &0.0067 &$-$22.55 &rh701938 & 24971&--\\
1144+35 &&11 47 22.1 &+35 01 07.3 &1 &1.81 &0.0630 &$-$23.40 &rh700862 &  3000&--\\
1217+29 &NGC4278&12 20 06.82 &+29 16 50.5 &12 &1.75 &0.0021 &$-$20.38 &rh701004a01 &  5810&--\\
1254+27 &NGC4839&12 57 24.37 &+27 29 52.7 &1 &0.90 &0.02464 &$-$23.48 &rp800009n00 & 19713& 7.5\\
1256+28 &NGC4869&12 59 21.3 &+27 54 40.3 &13 &0.91 &0.0224 &$-$22.13 &rh800242a04 & 36932& 6.9\\
1257+28 &NGC4874&12 59 35.7 &+27 57 33 &1 &0.91 &0.0239 &$-$23.20 &rh800242a04 & 36932&--\\
1317+33 &NGC5098&13 20 14.72 &+33 08 36.0 &1 &1.03 &0.0379 &$-$23.30 &rh800696n00 & 18908&--\\
1318+34 &&13 20 35.3 &+34 08 22 &1 &0.99 &0.0232 &$-$22.32 &rh704040n00 & 22510&--\\
1321+31 &NGC5127&13 23 45.00 &+31 33 55.7 &4 &1.16 &0.0161 &$-$22.42 &rh701829a01 & 21781&--\\
1322+36 &NGC5141&13 24 51.43 &+36 22 42.8 &1 &0.96 &0.0175 &$-$22.51 &rh702076n00 & 35078&--\\
1346+26 &A1795&13 48 52.6 &+26 35 35 $^*$&1 &1.12 &0.0633 &$-$24.03 &rh800222a01 & 10966&--\\
1350+31 &3C293&13 52 17.83 &+31 26 46.3 &14 &1.29 &0.0452 &$-$23.06 &rh700366 &  6136&--\\
1422+26 &&14 24 40.51 &+26 37 30.5 &1 &1.61 &0.0370 &$-$22.63 &rh701830 & 31329&--\\
1525+29 &A2079&15 27 44.40 &+28 55 6.5 &15 &2.38 &0.0653 &$-$24.20 &--&--&--\\
1553+24 &&15 56 03.90 &+24 26 52.9 &1 &4.42 &0.0426 &$-$23.13 &rh701565 & 39080&--\\
1610+29 &NGC6086&16 12 36.7 &+29 29 16 $^*$&1 &3.10 &0.0313 &$-$22.90 &rh800650n00 & 45547&--\\
1615+35 &NGC6109&16 17 40.50 &+35 00 14.7 &3 &1.47 &0.0296 &$-$22.85 &rh800164n00 & 33399&--\\
1621+38 &NGC6137&16 23 03.11 &+37 55 20.2 &1 &1.09 &0.031 &$-$23.60 &rh800165n00 & 31560&--\\
1626+39 &3C338&16 28 38.22 &+39 33 04.1 &1 &0.88 &0.0303 &$-$23.80 &rh800429a01 & 20755&--\\
1833+32 &3C382&18 35 03 &+32 41 47 &1 &7.28 &0.0586 &$-$24.20 &rh700368 &  4943&--\\
1855+37 &&18 57 37.5 &+38 00 33 $^*$&1 &8.01 &0.0552 &$-$24.50 &rp701445n00 &  8771&--\\
2116+26 &&21 18 33.0 &+26 26 49.4 &1 &9.65 &0.0164 &$-$22.59 &--&--&--\\
2229+39 &3C449&22 31 20.575 &+39 21 29.76 &8 &11.05 &0.0171 &$-$22.18 &rh704035n00 & 19064&--\\
2236+35 &&22 38 29.41 &+35 19 47.0 &1 &8.60 &0.0277 &$-$22.70 &rh702075 & 30971&--\\
2335+26 &3C465&23 35 58.97 &+26 45 16.16 &9 &4.91 &0.0301 &$-$23.85 &rh800715 & 61933&--\\
\end{tabular}
\end{table*}

\newpage

\begin{table*}
\caption{Results of model fits. Sources with too few counts to perform
a radial fitting, or those where the more complicated models show no
improvement over the simple point-source model, are not given. Sources
where the $\beta$+point model fails an F-test for improvement over a
$\beta$ model at the 95 per cent confidence level are denoted with a *
. If these sources then have a small value for the core radius in the
beta model, all the counts are assigned to the point source, otherwise
the $\beta$+point counts are used as a best estimate of the central
point-like emission.}
\label{models}
\begin{tabular}{l@{\hspace{1cm}}ll@{\hspace{1.5cm}}llll@{\hspace{1.5cm}}llll}
B2 & \multicolumn {2} {l} {Point  source}  & \multicolumn {4} {l} {$\beta$ model}  & \multicolumn {4} {l} {$\beta$+point model} \\
name    & $\chi^{2}$ &  df      & $\beta$ & $\theta_{core}$ & $\chi^{2}$ & df &  $\beta$ & $\theta_{core}$ & $\chi^{2}$ & df   \\
&     && & (arcsec) && & & (arcsec) & & \\ 
\hline
0055+26 &9780 &11 &0.9 &254 &43 &10 &0.9 &267 &9.4 &9 \\
0326+39 &1474 &41 &0.35 &0.05 &70 &39 &0.35 &60 &44 &38 \\
1040+31 &2173 &41 &0.35 &0.05 &42 &39 &0.35 &28 &34 &38 \\
1122+39 &707 &10 &0.67 &20 &4.8 &9 &0.75 &25 &4.6 $^*$&8 \\
1217+29 &8.4 &9 &0.67 &1.0 &4.3 &8 &0.9 &10 &2.3 &7 \\
1254+27 &22 &9 &0.5 &5.0 &4.3 &9 &0.9 &37.0 &3.8 $^*$&8 \\
1257+28 &522 &7 &0.5 &102 &6.0 &6 &0.5 &290 &0.38 &5 \\
1317+33 &2464 &30 &0.5 &8 &27 &29 &0.5 &10 &25 $^*$&28 \\
1346+26 &7734 &31 &0.67 &46 &303 &30 &0.5 &28 &48 &29 \\
1422+26 &1300 &10 &0.5 &33 &7.6 &9 &0.5 &47 &4.7 $^*$&8 \\
1615+35 &4.1 &5 &0.5 &$<$1.0 &1.7 &4 &0.75 &69 &0.069 &3 \\
1621+38 &330 &9 &0.5 &6 &11 &8 &0.5 &12 &8.2 $^*$&7 \\
1626+39 &39840 &96 &0.3 &7 &77 &95 &0.3 &9 &73 &94 \\
1855+37 &6714 &41 &0.35 &35 &54 &39 &0.4 &70 &32 &38 \\
2229+39 &4102 &11 &0.9 &275 &14 &10 &0.9 &282 &4.0 &9 \\
2236+35 &89 &11 &0.67 &4.8 &11 &10 &0.75 &10 &8.4 $^*$&9 \\
\end{tabular}
\end{table*}

\newpage

\begin{table*}
\caption{Net counts within the given source radius and the estimated contribution from unresolved emission (point counts). Sources marked with
an asterisk lie in cluster environments, so the net counts are more
dependent on the radii chosen for source and background regions than
for field galaxies. Details of any foreign sources excluded from these
regions are not shown. Consult Table \ref{sample} to see if counts are
for the HRI or PSPC. All PSPC counts are from the range 0.2-1.9 keV
only, where the PSF is well modelled. Note: For B2 0055+26, counts and
radii quoted are for position angles 125-290 degrees only, due to the
asymmetry of the source. `Bkg radii' shows the inner and outer
background radii.}
\label{counts}
\begin{tabular}{@{}lllllclll@{}}
 B2 & Net  &  & Point model  &  & Percentage of counts& Source radius & Bkg radii\\
name & counts & Error &  counts & Error &in point source & (arcsec) & (arcsec)\\
\hline
0055+30 &284&22&284&22&100&50 &50 &256 \\
0055+26 &1066&42&60&10&  6&120 &120 &256 \\
0104+32 $^*$&165&17&165&17&100&30 &60 &256 \\
0120+33 $^*$&245&27&$<$245&--&--&30 &45 &60 \\
0149+35 $^*$&207&23&$<$207&--&--&20 &20 &35 \\
0206+35 $^*$&63&10&63&10&100&15 &60 &256 \\
0222+36 &32&14&$<$32&--&--&30 &50 &128 \\
0258+35 &20.9&9.6&$<$20.9&--&--\\
0326+39 &638&56&134&14& 21&180 &180 &340 \\
0331+39 &462&39&462&39&100&100 &100 &500 \\
0722+30 &$<$69&--&$<$69&--&--\\
0755+37 &177&19&177&19&100&45 &45 &256 \\
0800+24 &$<$10&--&$<$10&--&--\\
0836+29A $^*$&$<$8&--&$<$8&--&--\\
0924+30 &75&29&$<$75&--&--&32 &32 &36 \\
1040+31 &942&82&139&17& 15&180 &180 &340 \\
1113+29 $^*$&26&15&26&15&100&75 &75 &256 \\
1122+39 &159&22&4.1&6.8&  3&50 &100 &256 \\
1144+35 &71&11&71&11&100&50 &50 &256 \\
1217+29 &104&14&86&14& 82&50 &50 &256 \\
1254+27 $^*$&267&71&236&55& 88&50 &50 &120 \\
1256+28 $^*$&$<$12&--&$<$12&--&--\\
1257+28 $^*$&219&44&34&14& 15&45 &60 &120 \\
1317+33 &531&31&20&12&  4&60 &60 &256 \\
1318+34 &11&11&$<$11&--&--&30 &128 &256 \\
1321+31 $^*$&48&20&$<$48&--&--&50 &60 &128 \\
1322+36 &54&27&$<$54&--&--&50 &50 &90 \\
1346+26 $^*$&11590&150&61&19&  1&256 &256 &360 \\
1350+31 &20.7&9.5&$<$20.7&--&--&50 &50 &256 \\
1422+26 &254&42&18.8&6.6&  7&100 &200 &256 \\
1553+24 &122&24&122&24&100&50 &50 &256 \\
1610+29 $^*$&69&17&69&17&100&30 &100 &256 \\
1615+35 &102&22&62&11& 61&50 &50 &256 \\
1621+38 &440&29&77&15& 18&50 &50 &256 \\
1626+39 $^*$&3516&77&59&22&  2&45 &60 &120 \\
1833+32 &2913&56&2913&56&100&45 &60 &120 \\
1855+37 &1191&83&75&13&  6&180 &180 &340 \\
2229+39 &301&56&24.5&7.6&  8&150 &325 &375 \\
2236+35 &160&34&40&12& 25&75 &75 &250 \\
2335+26 $^*$&439&40&439&40&100&60 &100 &256 \\
\end{tabular}
\end{table*}

\newpage

\begin{table*}
\caption{Radio and X-ray core flux densities and luminosities. 
For the radio core, references are as follows: 1, Giovannini et. al (1990); 2,
Fanti et al. (1987); 3, Venturi et al. (1993); 4, Fomalont, private communication; 5,
various, see text; 6, measured from maps supplied by R. Morganti. A
source of type U is undetected in the X-ray, and the X-ray measurement
is a $3\sigma$ upper limit. In a source of type P a compact source is
detected in the X-ray, and all the X-ray counts are attributed to the
X-ray core. In type R, a mixture of extended and compact X-ray
emission is found. The X-ray core value is from the best-fit
point-source contribution to a multi-component model. Sources of type
L are upper limits corresponding to the total detected counts on a
source, either because there are too few counts to perform an adequate
radial profile fit or because the source has complicated structure
which is not well characterised by our models. $L\rm ^{0.2-2.5
keV}_{point}$ is the estimated luminosity of the X-ray point-like component
between 0.2 and 2.5 keV. The X-ray values assume a power-law spectrum with energy index 0.8, and errors in the 1 keV luminosity density are statistical only.}
\label{lums}
\begin{tabular}{@{}lllllllll@{}}
B2 & $S\rm ^{5GHz}_{core}$ & Ref.& $L\rm ^{5GHz}_{core}$ & $S\rm ^{1 keV}_{point}$  & $L\rm ^{1 keV}_{point}$ & $L\rm ^{0.2-2.5}$ & Type\\

name &  (mJy) &  & ($10^{20}$ W Hz$^{-1}$ sr$^{-1}$) & (nJ) &
($10^{15}$ W Hz$^{-1}$ sr$^{-1}$) & (10$^{35}$ W) \\
\hline
0055+30 &450 &3 &430 &190 &$18 \pm1 $&1.3 &P\\
0055+26 &8.6 &4 &66 &11 &$8.5 \pm1 $&0.62 &R\\
0104+32 &92 &1 &90 &64 &$6.3 \pm0.7 $&0.46 &P\\
0120+33 &1.4 &1 &1.3 &$<$83 &$<$7.8 &$<$0.56 &L\\
0149+35 &5 &1 &4 &$<$59 &$<$5.2 &$<$0.38 &L\\
0206+35 &106 &1 &510 &29 &$14 \pm2 $&1.0 &P\\
0222+36 &90 &1 &330 &$<$10 &$<$4 &$<$0.3 &L\\
0258+35 &$<$15 &1 &$<$13 &$<$10 &$<$1 &$<$0.08 &L\\
0326+39 &70 &1 &140 &36 &$7.3 \pm0.8 $&0.53 &R\\
0331+39 &125 &1 &175 &230 &$33 \pm3 $&2.4 &P\\
0722+30 &51 &1 &64 &$<$20 &$<$3 &$<$0.2 &U\\
0755+37 &190 &1 &1110 &85 &$51 \pm5 $&3.7 &P\\
0800+24 &3 &1 &20 &$<$3 &$<$2 &$<$0.1 &U\\
0836+29A &8.2 &1 &120 &$<$4 &$<$6 &$<$0.4 &U\\
0924+30 &$<$0.4 &1 &$<$1.00 &$<$20 &$<$5 &$<$0.3 &L\\
1040+31 &55 &1 &240 &16 &$7.5 \pm0.9 $&0.55 &R\\
1113+29 &41 &1 &340 &10 &$8 \pm5 $&0.6 &P\\
1122+39 &6 &1 &0.9 &1 &$0.02 \pm0.03 $&0.001 &R\\
1144+35 &250 &1 &3400 &170 &$240 \pm40 $&18 &P\\
1217+29 &350 &1 &5.28 &100 &$0.16 \pm0.03 $&0.011 &R\\
1254+27 &2.3 &5 &4.8 &25 &$5.2 \pm1 $&0.38 &R\\
1256+28 &2 &1 &3 &$<$2 &$<$0.3 &$<$0.03 &U\\
1257+28 &1.1 &1 &2.2 &6 &$1 \pm0.4 $&0.08 &R\\
1317+33 &8 &6 &40 &7 &$3 \pm2 $&0.2 &R\\
1318+34 &84 &5 &150 &$<$3 &$<$0.6 &$<$0.04 &L\\
1321+31 &21 &1 &19 &$<$10 &$<$1 &$<$0.09 &L\\
1322+36 &75 &1 &79 &$<$10 &$<$1 &$<$0.07 &L\\
1346+26 &53 &1 &730 &36 &$52 \pm20 $&3.7 &R\\
1350+31 &100 &1 &700 &$<$20 &$<$20 &$<$1 &L\\
1422+26 &25 &1 &120 &4 &$2 \pm0.7 $&0.1 &R\\
1553+24 &53.6 &2 &333 &11 &$6.8 \pm1 $&0.49 &P\\
1610+29 &$<$6 &1 &$<$20 &12 &$4.3 \pm1 $&0.31 &P\\
1615+35 &28 &1 &84 &13 &$3.9 \pm0.7 $&0.28 &R\\
1621+38 &50 &1 &160 &16 &$5.3 \pm1 $&0.38 &R\\
1626+39 &100 &1 &314 &20 &$6 \pm2 $&0.4 &R\\
1833+32 &190 &1 &2240 &6090 &$7500 \pm100 $&543 &P\\
1855+37 &$<$100 &1 &$<$1040 &35 &$38 \pm7 $&2.8 &R\\
2229+39 &37 &1 &37 &15 &$1.5 \pm0.5 $&0.11 &R\\
2236+35 &8 &1 &20 &14 &$3.7 \pm1 $&0.27 &R\\
2335+26 &230 &1 &713 &66 &$21 \pm2 $&1.5 &P\\
\end{tabular}
\end{table*}

\newpage

\begin{figure*}
\begin{center}
\leavevmode
\epsfxsize 15cm
\epsfbox{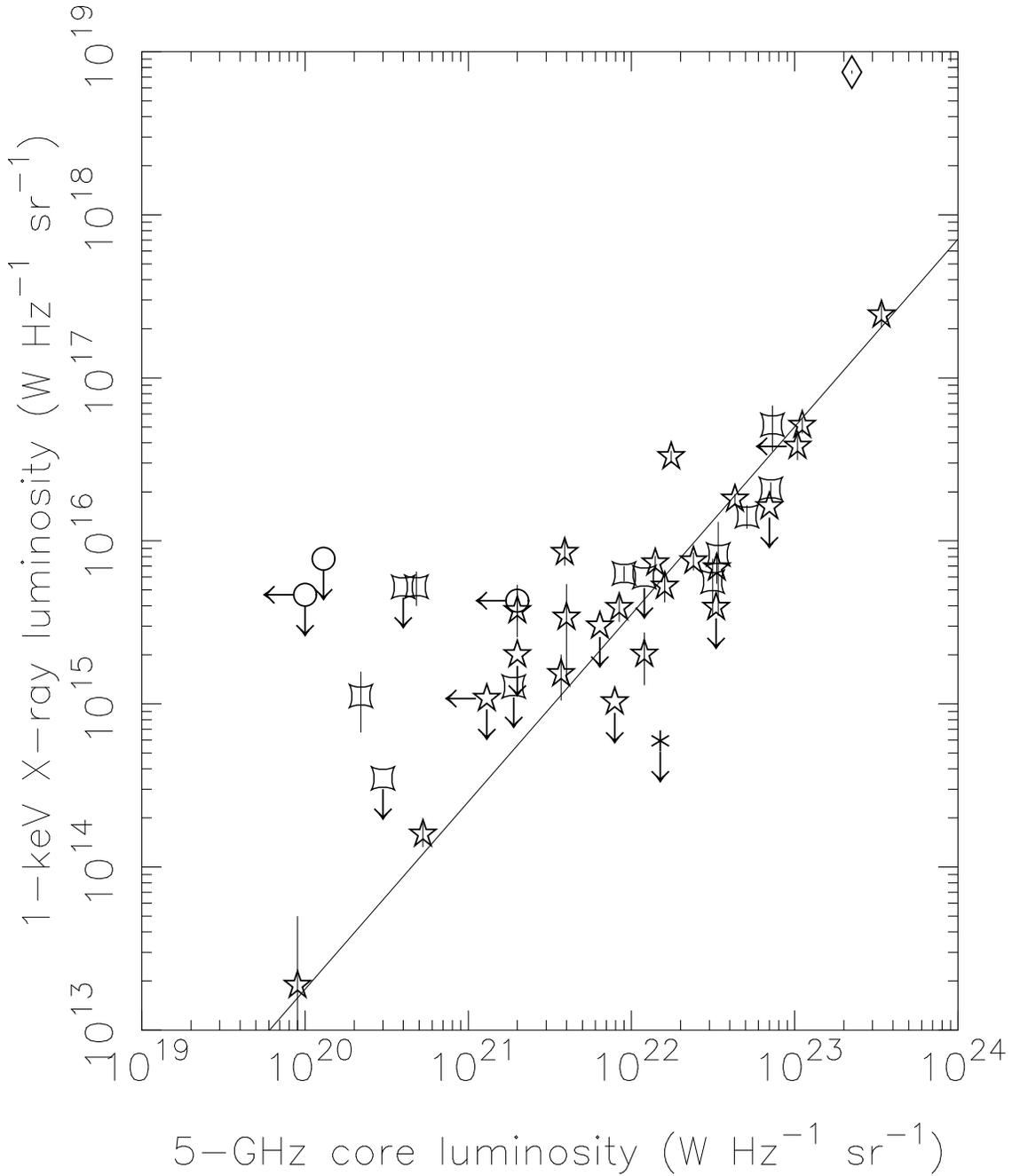}
\end{center}
\caption{X-ray core luminosity is plotted vs radio core
luminosity. The line drawn is derived from the Schmitt regression
[(excluding a broad-line radio galaxy and a starburst galaxy
(see text)] and has a slope of 1.15 and an intercept of -9.82.
  Cluster sources are shown by
squashed squares, an open circle denotes a relic source, a cross
represents a starburst galaxy, and a closed star is everything
else. The excluded broad line FRII source 3C382 has been displayed as
a diamond; it is over-bright in X-rays for its core-radio strength,
suggesting an additional X-ray emission component, consistent with
unified models. The arrows show where upper limits have been taken and
in all other cases 1$\sigma$ error bars are shown.  }
\label{xrayradio1}
\end{figure*}

\newpage

\begin{figure*}
\begin{center}
\leavevmode
\epsfxsize 15cm
\epsfbox{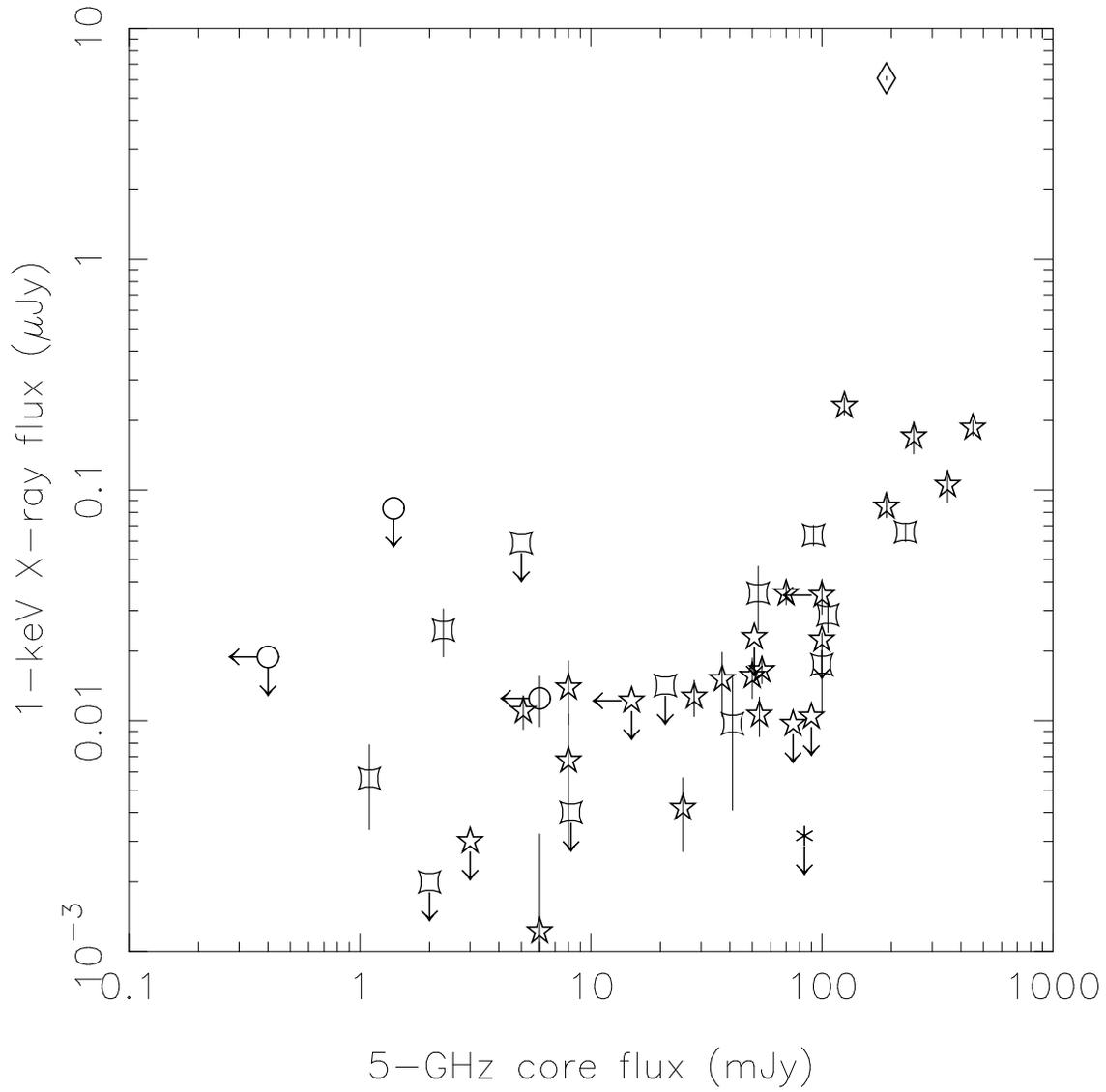}
\end{center}
\caption{X-ray against radio flux density. Symbols as for Fig. \ref{xrayradio1}. The correlation persists here, and is significant at the $>$99.99 level.}
\label{xrayradio2}
\end{figure*}

\end{document}